\documentstyle[epsf,aps]{revtex} 
\title{\bf Quantum structure of the
non-Abelian Weizs\"{a}cker-Williams field for a very large nucleus.}

\author{Yuri V. Kovchegov \footnote{Electronic address:
yuri@phys.columbia.edu}}

\begin{document}
\maketitle
\begin{center}
 \it{Department of Physics, Columbia University, New York, NY 10027,
USA}
 
\end{center}

\begin{abstract}

   We consider the McLerran-Venugopalan model for calculation of the
small-$x$ part of the gluon distribution function for a very large
ultrarelativistic nucleus at weak coupling. We construct the Feynman
diagrams which correspond to the classical Weizs\"{a}cker-Williams
field found previously [Yu. V. Kovchegov, Phys. Rev. D {\bf 54}, 5463
(1996)] as a solution of the classical equations of motion for the
gluon field in the light-cone gauge. Analyzing these diagrams we
obtain a limit for the McLerran-Venugopalan model. We show that as
long as this limit is not violated a classical field can be used for
calculation of scattering amplitudes.

\ \\ PACS number(s): 12.38.Bx, 12.38.Aw, 24.85.+p

\end{abstract}

%\twocolumn

\section{Introduction}

    An interesting problem in nuclear and particle physics is
computing gluon distribution functions for a nucleus at small values
of Bjorken $x$. Some time ago the problem was attacked by L. McLerran
and R. Venugopalan \cite{Larry}. In their model they consider a very
large nucleus, larger than a physical nucleus, which is moving
ultra-relativistically and effectively looks like a pancake in the
transverse plane. In that plane the nucleus is described by a
classical color charge density $\rho(\underline{x})$ . The strong
coupling constant $\alpha_s$ is small, which gives a lower limit on
the typical scale of the transverse momentum in the problem: $k_\perp
\gg \Lambda_{\mbox{QCD}}$. Actually, to apply successfully the
perturbation theory one also has to satisfy another condition:
$k_\perp > \alpha_s \mu$ \cite{Larry}. It was shown that the relevant
transverse coordinate scale in a scattering process is small, but it
should not be too small \cite{Larry,Alex} : $k_\perp \ll \mu$, where
$\mu$ is the typical scale of the color charge density
fluctuations. In \cite{Larry} it was assumed that one has to find the
classical gluon field in the light-cone gauge, treating the nucleus as
a classical source, and that this field will dominate in the
distribution function. Quantum effects will come in as virtual
corrections. For this approximation to be valid one needs this
$k_\perp \ll \mu$ condition.
   
     Since the nucleus is ultrarelativistic and Lorentz-contracted to
almost a plane, a small-$x$ gluon in the nucleus ``sees'' not just one
nucleon in the longitudinal direction, but in the order of $A^{1/3}$
of them, with $A$ the atomic number. That is an essential feature of
the model at hand --- longitudinal coherence of the nucleus. In order
to find an average value of any observable with longitudinal coherence
length long compared to the nucleus, one has to calculate this
observable for a given color charge density $\rho (\underline{x})$ and
then average it over all $\rho$ with the Gaussian measure \cite{me}.
   
     The correct classical gluon field, as a solution of the classical
non-Abelian equations of motion, has recently been found
\cite{me,Jamal}. An important issue is the way one has to treat the
nucleus. The ultrarelativistic nucleus is a source of color charge in
the classical Yang-Mills equations of motion. Until recently it was
treated just as an infinitely thin sheet lying in the transverse plane
--- a delta function along the light cone \cite{Larry}.  This
approximation happened to be not quite accurate, and leads to infrared
problems \cite{Raju}. Later, a solution for the gluon field has been
constructed which incorporates the effects of a finite size of the
nucleus in the longitudinal direction \cite{me,Jamal}. Our solution
\cite{me} and the one found in \cite{Jamal} by J. Jalilian-Marian et
al. are equivalent, they give the same expression for the gluon
distribution function $\left< A_i^a (\underline{x}) A_i^a
(\underline{y}) \right>$.

      In our approach \cite{me} we formulate the McLerran-Venugopalan
model in terms of point charges: each ``nucleon'' was taken, for
simplicity of color algebra, to be a quark-antiquark pair. These
valence quarks and antiquarks were free to move inside the nucleons
(spheres of equal radius in the rest frame), but unable to get
out. Finding the solution for the gluon field in covariant gauge, we
then performed a gauge transformation to the light-cone gauge and
obtained the non-Abelian Weizs\"{a}cker-Williams field for the
ultrarelativistic nucleus (see Eq. (10) in \cite{me} ):

\begin{eqnarray}
{\underline {A}} (\underline {x}, x_{-} ) = { g \over {2 \pi }}
\sum_{a=1}^8 \sum_{i=1}^N \left( S(\underline {x}, x_{-i} ) T^a
(T_i^a) S^{-1} (\underline {x},
x_{-i}){{\underline{x}}-{\underline{x}}_i \over
|{\underline{x}}-{\underline{x}}_i|^2}\theta (x_{-} - x_{-i})
\right. \nonumber
\end{eqnarray}
\begin{eqnarray}
 - \left. S(\underline {x}, x'_{-i} ) T^a (T_i^a) S^{-1} (\underline
{x}, x'_{-i}){{\underline{x}}-{\underline{x}'}_i \over
|{\underline{x}}-{\underline{x}'}_i|^2}\theta (x_{-} -
x'_{-i})\right), A_+ = 0 , A_- = 0 .
\label{clsol}
\end{eqnarray}
Here ${\underline{x}}_i$ and ${\underline{x}'}_i$ are the transverse
coordinates of the quark and antiquark in the $i$th nucleon, $x_{-i}$
and $x'_{-i}$ are the light-cone coordinates, $N$ is the total number
of nucleons in the nucleus, $T^a$ are $SU(3)$ generators, $(T_i^a)$
are similar generators in the color space of each nucleon. The
classical current in a non-Abelian gauge theory is given by $j=T^a j^a
= T^a g {\overline q}_\alpha \gamma_{\mu} (T^a)_{\alpha \beta} q_\beta
$, so the matrix $(T^a)_{\alpha \beta}$ can be understood as a part of
the coupling. It is a matrix in the color space of a nucleon, which is
different from the color space of $T^a$. These two matrices act in the
different color spaces, and, therefore, commute. The non-Abelian
Weizs\"{a}cker-Williams field (\ref{clsol}) is used in calculation of
such quantities as the gluon distribution function. Therefore, the
condition that the initial and final states of the nucleons should be
color singlets is imposed on a product of two fields, but not on the
field itself.

   $S(\underline {x},x_{-})$ is a matrix which effects the gauge
transformation from covariant to the light-cone gauge, and is given by
(Eq. (18) in \cite{me})
\begin{eqnarray}
S(\underline {x}, x_{-} ) = \mbox{P} \exp \left( { -ig
\int_{-\infty}^{x_{-}} dx'_{-} A'_{+} (\underline {x}, x'_{-})}
\right) = \prod_{i=1}^N \exp \left[ {i g^2 \over {2 \pi }}
\sum_{a=1}^8 T^a (T_i^a) \ln \left({|\underline {x} - {\underline
{x}}_i| \over { |\underline {x} - {\underline {x}}'_i|}} \right)
\theta (x_{-} - x_{-i})\right].
\label{cutoff}
\end{eqnarray}
Here $A'_{+} (\underline {x}, x'_{-})$ is the gluon field in the
covariant gauge and the nucleons are labeled according to their
positions along the $x_-$-axis, i.e. , the greater the $x_-$
coordinate of a nucleon, the greater is its label $i$. In
Eq. (\ref{cutoff}) we neglect the contribution of the ``last''
nucleon, i.e., the nucleon (or several nucleons) whose quarks or
antiquarks may overlap the point $x_{-}$ at which we calculate
$S(\underline {x}, x_{-} )$. This is justified, because if the
nucleons are ordered in longitudinal direction there is only one such
nucleon. The exponential in Eq. (\ref{cutoff}) corresponding to this
``last'' nucleon gets cancelled by color algebra once we try to
calculate the field in Eq. (\ref{clsol}).  If there are several
``last'' nucleons, then we can just throw them away, since the nucleus
is considered to be large and the contribution of a few of its
nucleons is not substantial.

     The choice of $S(\underline {x},x_{-})$ in Eq. (\ref{cutoff}) to
be a path-ordered integral from $-\infty$ to $x_-$ is not unique. One
could also take a path-ordered integral from $x_-$ to $+\infty$, or
construct some other expression which would enable us to perform the
desired gauge transformation.

     In this paper we will try to understand the quantum structure of
the classical field given by Eq. (\ref{clsol}). We shall show that
this field corresponds to a particular set of Feynman diagrams in the
light-cone gauge. Expanding the right-hand side of Eq. (\ref{clsol})
in powers of $g$, we start by giving the Feynman diagrams
corresponding to the non-Abelian Weizs\"{a}cker-Williams field at
lowest orders in the coupling constant. In Sect. II we will present
and calculate the diagrams corresponding to the classical field at
orders $g$ and $g^3$ for two nucleons in the nucleus. An easy and
elegant way to sum the diagrams at order $g^3$ and higher orders is by
applying the Ward identity \cite{t'Hooft,Sterman}. We will briefly
review this technique for the light-cone gauge.

    In Sect. III we will write down and evaluate those diagrams giving
the order $g^5$ contribution to the classical gluon field of two
nucleons in the nucleus. At this level we shall see that taking the
color average in the color space of each nucleon, similar to what one
has to do to calculate the correlation function of two fields, is
crucial for the equivalence of the diagrams and the classical field,
as well as for calculating the field itself.  At higher orders ($g^7$
and above) the classical solution ceases to be a good approximation to
the physical gluon field of two nucleons, since quantum corrections
become important. That is, we find a limit to the classical approach,
which happens to be just two gluons per nucleon.

    We conclude in Sect. IV by constructing the lowest order diagrams
contributing to the scattering cross-section of the ultrarelativistic
nucleus on a heavy quarkonium. In this example we show that if one
limits exchanged gluons to two per nucleon, all the diagrams are
essentially ``classical'', that is this scattering is described by a
classical field. That shows that the classical approximation is valid
at this order and allows one to use it in the calculation of many
other processes such as charm production, etc.

\section{Lowest order diagrams}

    Our goal now is to find the Feynman diagrams in the light-cone
gauge giving the non-Abelian Weizs\"{a}cker-Williams field for a
nucleus. To understand the general structure of these diagrams we
consider a simple case of two nucleons in the nucleus. The
generalization to a large number of nucleons will be simple, once we
understand what kind of diagrams are needed to construct the classical
gluon field.

    We start with two nucleons, which are ordered and separated in the
longitudinal direction ($x_{-2} > x_{-1}$). Then, expanding
Eq. (\ref{clsol}) for $N=2$, we obtain the classical field of this
system at lowest order:
\begin{eqnarray*}
{\underline {A}}^a (\underline {x}, x_{-} ) = { g \over {2 \pi }}
(T^a_1) \left( {{\underline{x}}-{\underline{x}}_1 \over
|{\underline{x}}-{\underline{x}}_1|^2} \theta (x_{-} - x_{-1}) -
{{\underline{x}}-{\underline{x}'}_1 \over
|{\underline{x}}-{\underline{x}'}_1|^2} \theta (x_{-} - x'_{-1})
\right)
\end{eqnarray*}
\begin{eqnarray}
+ { g \over {2 \pi }} (T^a_2) \left(
{{\underline{x}}-{\underline{x}}_2 \over
|{\underline{x}}-{\underline{x}}_2|^2} \theta (x_{-} - x_{-2}) -
{{\underline{x}}-{\underline{x}'}_2 \over
|{\underline{x}}-{\underline{x}'}_2|^2} \theta (x_{-} - x'_{-2})
\right) + o (g^3).
\label{order1}
\end{eqnarray}

\begin{figure}
\begin{center}
\epsfxsize=5cm
\epsfysize=1cm
\leavevmode
\hbox{ \epsffile{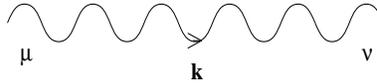}}
\end{center}
\caption{Gluon propagator in the light-cone gauge (see text).}
\label{prop}
\end{figure}

\begin{figure}
\begin{center}
\epsfxsize=5cm
\epsfysize=5cm
\leavevmode
\hbox{ \epsffile{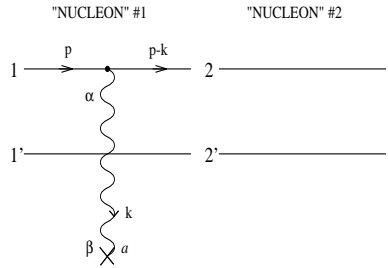}}
\end{center}
\caption{Diagram giving the classical field in the light-cone gauge at
lowest order.}
\label{or1}
\end{figure}

    Before discussing the diagrams giving this field (one of which is
shown in Fig. \ref{or1} ), we make a few comments about the way we
treat the gluon propagator in light-cone gauge, since it will be very
important in the calculations to follow. The gluon propagator in
light-cone gauge is given by:
\begin{eqnarray*}
P_{\mu \nu} (k) = - {i \over {k^2}} \left(g_{\mu \nu} - {\eta_\mu
k_\nu \over {k_+}} - {\eta_\nu k_\mu \over {k_+}}\right),
\end{eqnarray*}
where color indices have been suppressed and where $\eta$ is such that
for any four-vector $v$: $\eta \cdot v = v_+$. In calculating Feynman
diagrams one has to deal with the singularity of this propagator at
$k_+=0$. We regularize it in such a way that the propagator becomes
\begin{eqnarray}
P_{\mu \nu} (k) = - {i \over {{k}^2}} \left(g_{\mu \nu} - {\eta_\mu
k_\nu \over {k_+ -i\epsilon}} - {\eta_\nu k_\mu \over {k_+
+i\epsilon}}\right).
\end{eqnarray}
If the momentum $k$ in a term in the propagator flows from $\eta$ to
$k$ we use $-i\epsilon$ (If the momentum flows from $\mu$ to $\nu$, as
in Fig. \ref{prop}, then for a term like $\eta_\mu k_\nu \over {k_+}$
we say that it flows from $\eta$ to $k$.). If it flows the other way
we take $+i\epsilon$, where $\epsilon$ is some infinitesimal
number. This unusual choice of the $i\epsilon$ is necessary to
reproduce the classical solution (\ref{clsol}) from Feynman diagrams.

The Fourier transform of $1 \over {k_+ - i \epsilon}$ gives a
theta-function $\theta (x_-)$. In principle we could regularize the
propagator in other ways, for example by taking the $i \epsilon$ with
an opposite sign or by taking the principal value of the $k_+$
integral. The Fourier transform then would give $\theta(-x_-)$ or
$\epsilon(x_-)$. In that sense our choice of regularization is
arbitrary. It is done in the spirit of our choice of the matrix
responsible for the gauge transformation in \cite{me}. We want to
reproduce the field which was obtained using one particular choice of
that matrix [see Eq. (\ref{cutoff})], so we have to regularize the
propagator in a corresponding way.

     Now consider the diagram shown in Fig. \ref{or1}. The fermion
lines correspond to the quark and antiquark in the first and second
nucleons respectively. The cross at the end of gluon line denotes the
point where we measure the gluon field. The incoming and outgoing
quark lines are on-shell, their momenta are almost identical and in
light-cone coordinates are given by $p_\mu \approx (p_+, 0 ,
{\underline{0}})$.

    Each nucleon in our model is a bound state of a quark-antiquark
pair. The state has a unit normalization. The quarks in the nucleons
are not very far off-shell, which allows us to treat them as on-shell
incoming and outgoing particles in our calculation. The total
transverse momentum of gluons interacting with a nucleon is small
compared to the typical momentum in the nucleon's wave function. This
results from the fact that the total transverse momentum of the gluons
is cut off by the inverse size of the nucleus, which is much larger
then the size of the nucleons. So, the wave function of the final
state of a nucleon is approximately the same as the initial state wave
function and does not depend much on the total transverse momentum of
the gluons coming into the nucleon, since it is small. That means that
the product of these wave functions is just a square of the initial
state wave function, which gives us just a factor of one after
momentum integration due to the normalization of the bound state. For
that reason we are not going to explicitly include the wave function
in our calculations. In the calculations we make in this and the
following sections to find the classical field we are not computing an
amplitude of a physical process. Therefore, we do not require the
initial and final states of the nucleons to be color singlets unless
specified separately. Also we do not impose any limit on the magnitude
of the gluon's transverse momentum. In a physical process, such as
scattering, the total transverse momentum of the gluons interacting
with a nucleon is cut off by the inverse size of the nucleus. However,
this does not limit the transverse momentum of each individual
gluon. The only possible cutoff on that momentum is the inverse size
of a nucleon, but it is very large. That allows us to integrate the
transverse momentum up to infinity.

 Using the formula for the gluon propagator in the light-cone gauge,
we can write down the contribution of the graph in Fig. \ref{or1} as:
\begin{eqnarray}
 - { i \over {k^2}}{\left( g_{\alpha \beta} - {\eta_\alpha k_\beta
\over {k_{+} - i \epsilon}} - {\eta_\beta k_\alpha \over {k_{+} + i
\epsilon}} \right)} i g {1 \over {2 p_+}} \tilde{u} (p-k)
\gamma_\alpha u(p) (T^a_1) (2 \pi) \delta (k_-) = g {k_\beta^\perp
\over {{\underline{k}}^2}} {1 \over {k_{+} - i \epsilon}} (T^a_1) (2
\pi) \delta (k_-).
\label{lo}
\end{eqnarray}
The $\eta_\beta k_\alpha \over {k_{+} + i \epsilon}$ part of the
propagator gives $\tilde{u} (p-k) \gamma \cdot k u(p) = 0$ and ,
therefore, vanishes.  When $\beta=+$ the propagator is proportional to
$g_{\alpha +} - \eta_\alpha = 0$. When $\beta=-$ the amplitude is
again zero because $\tilde{u} (p-k) \gamma^- u(p) = 0 $, since the
transverse momentum of the quark is $p_\mu^\perp \approx 0$ (for
example see Appendix A of \cite{BL}). The only non-vanishing
contribution comes from $\beta = \perp$. But even in that case the
covariant part of the propagator ($g_{\alpha \beta}$) goes away. This
way we are left with the expression given on the right of
Eq. (\ref{lo}).  The factor of $(2 \pi) \delta (k_-)$ comes from the
condition that the outgoing quark line is almost on-shell. Formula
(\ref{lo}) is similar to the light-cone potential of a point charge
\cite{Mueller}. It has the same normalization except for a factor of
$(2 \pi)^2$ resulting from a prefactor in the Fourier
transform. Performing a Fourier transform of Eq. (\ref{lo}) in the
transverse and longitudinal directions we end up with
\begin{eqnarray}
g (T^a_1) \int {d^2 \underline{k} d k_+ dk_- \over { (2 \pi)^4}} e^{i
k_+ (x_- - x_{-1}) + i k_- (x_+ - x_{+1}) - i \underline{k} \cdot
({\underline{x}} - {\underline{x}}_1) } {\underline{k} \over
{{\underline{k}}^2}} {1 \over {k_{+} - i \epsilon}} (2 \pi) \delta
(k_-) = { g \over {2 \pi }} (T^a_1) {{\underline{x}}-{\underline{x}}_1
\over |{\underline{x}}-{\underline{x}}_1|^2}\theta (x_{-} - x_{-1}),
\end{eqnarray}
which looks exactly like the lowest order classical field emitted due
to one parton. Summing over the diagrams with the gluon line hooking
to each one of the four fermion lines gives the expression in
Eq. (\ref{order1}). A minus sign appears when the gluon is connected
to an antiquark line. This establishes the correspondence between the
classical field and the Feynman diagrams at lowest order in $g$.

    Let us try to go further and find the diagrams giving the field at
order $g^3$. First one has to write down the classical fields at this
order of the coupling, which is easily done by expanding
Eq. (\ref{clsol}) to the next order in $g^2$:
\begin{eqnarray*}
{\underline {A}^a} (\underline {x}, x_{-} ) = o(g) - { g^3 \over {(2
\pi)^2 }} \sum_{b,c=1}^8 f^{abc} (T_2^c) (T_1^b) 
\end{eqnarray*}
\begin{eqnarray}
\times \ln \left( {|{\underline{x}}-{\underline{x}}_1 | \over
|{\underline{x}}-{\underline{x}'}_1|} \right) \left(
{{\underline{x}}-{\underline{x}}_2 \over
|{\underline{x}}-{\underline{x}}_2|^2}\theta (x_{-} - x_{-2}) -
{{\underline{x}}-{\underline{x}'}_2 \over
|{\underline{x}}-{\underline{x}'}_i|^2}\theta (x_{-} - x'_{-2})\right)
+ o(g^5).
\label{f3}
\end{eqnarray}

\begin{figure}
\begin{center}
\epsfxsize=15cm
\epsfysize=6cm
\leavevmode
\hbox{ \epsffile{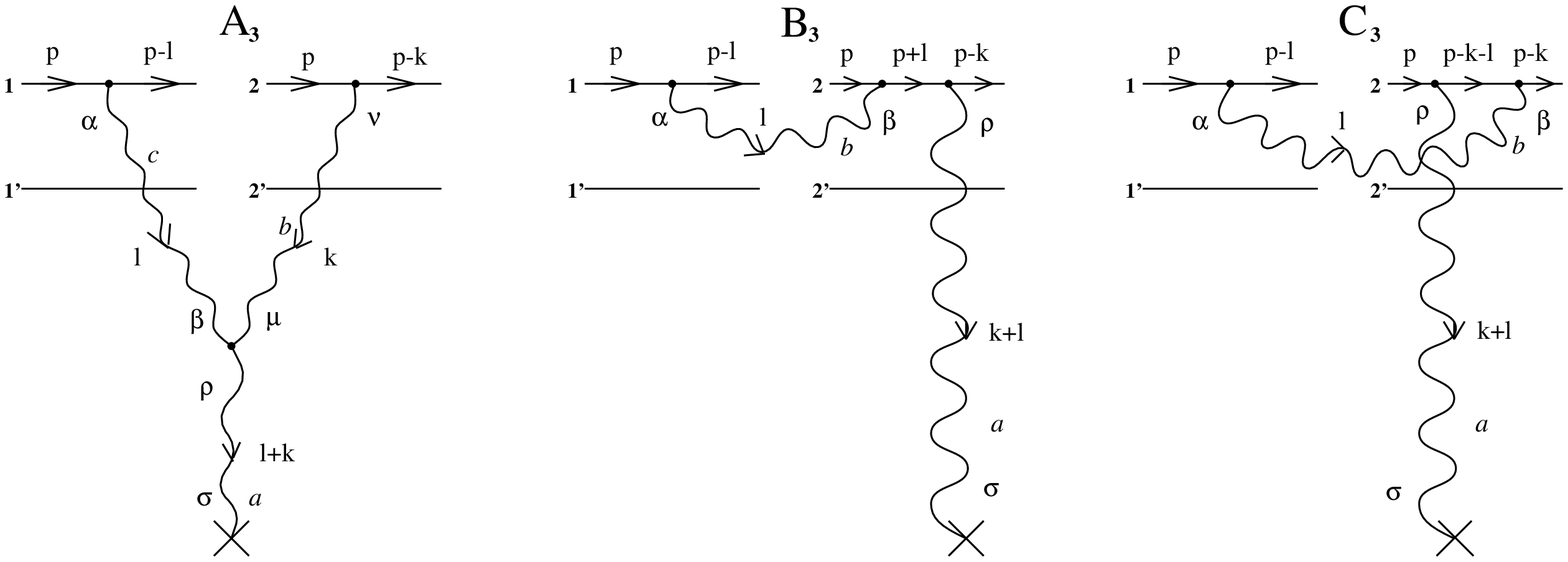}}
\end{center}
\caption{Diagrams giving the classical field in the light-cone gauge
at order $g^3$. The intersection of two gluon lines in C is not a
vertex.}
\label{or3}
\end{figure}

The claim is that in the light-cone gauge the sum of the diagrams in
Fig. \ref{or3}, together with all permutations (gluons connecting to
different pairs of quarks, each of them being in a different nucleon,
not just to 1 \& 2 like in the Fig. \ref{or3}, but also to 1 \& 2', 1'
\& 2, 1' \& 2' ) gives us the contribution to the classical field
presented in Eq. (\ref{f3}).

A brute force calculation yields, for the $\sigma = \perp$ component,
\begin{mathletters}
\begin{eqnarray*}
A_3 = - i g^3 \sum_{b,c=1}^8 f^{abc} (T_2^c) (T_1^b) \left(
{k_\sigma^\perp + l_\sigma^\perp \over
{\underline{k}^2(\underline{k}+\underline{l})^2}} {1 \over {k_+ - i
\epsilon}} {1 \over {k_+ + l_+ - i \epsilon}} + {k_\sigma^\perp +
l_\sigma^\perp \over {\underline{k}^2 \underline{l}^2}} {1 \over {l_+
- i \epsilon}}{1 \over {k_+ + l_+ - i \epsilon}} \right.
\end{eqnarray*}
\begin{eqnarray}
\left. -{k_\sigma^\perp + l_\sigma^\perp \over {\underline{l}^2
(\underline{k}+\underline{l})^2}} {1 \over {k_+ + l_+ - i \epsilon}}{1
\over {l_+ - i \epsilon}} - {l_\sigma^\perp \over {\underline{k}^2
\underline{l}^2}} {1 \over {k_+ - i \epsilon}}{1 \over {l_+ - i
\epsilon}} \right) (2 \pi)^2 \delta (k_-) \delta (l_-),
\end{eqnarray}
\begin{eqnarray}
B_3 + C_3 = - i g^3 \sum_{b,c=1}^8 f^{abc} (T_2^c) (T_1^b)
{k_\sigma^\perp + l_\sigma^\perp \over {\underline{l}^2
(\underline{k}+\underline{l})^2}} {1 \over {k_+ + l_+ - i \epsilon}}{1
\over {l_+ - i \epsilon}}(2 \pi)^2 \delta (k_-) \delta (l_-).
\end{eqnarray}
\end{mathletters}

In the calculation of the graphs $B_3$ and $C_3$ we take only the part
of the gluon propagator for the $l$-line which is longitudinally
polarized at the $\beta$-end of the line. The $\eta_\beta l_\alpha
\over {l_{+} + i \epsilon}$ part of the propagator gives $\tilde{u}
(p-l) \gamma \cdot l u(p) = 0$. The covariant part of the propagator,
i.e., the part proportional to $g_{\alpha \beta}$, is small. The
reason for that is quite straightforward. Suppose we have a gluon line
connecting two fermions which are separated by some distance $x_- > 0$
in the longitudinal direction. The typical $x_-$ is much larger than
the longitudinal size of the nucleons. If a gluon had a mass the
interaction described by the covariant part of the propagator would be
a short range interaction and would be suppressed. But in our case the
role of the mass is played by the transverse momentum of the gluon. We
take the covariant part of the gluon's propagator and perform a
Fourier transform along the $l_+$ direction. To localize the fermions
we take them to have some mass. In the infinite momentum frame, for a
fermion with non-zero mass $m$, its momentum is given by $p_\mu
\approx (p_+, {m^2 \over {2 p_+}} , {\underline{0}})$. Using the
condition that the fermion, after emitting a gluon, remains on-shell
($(p-l)^2 = m^2$) in the Fourier transform we obtain
\begin{eqnarray}
\int_{-\infty}^{+\infty} {d l_+ \over { 2 \pi}} {e^{i l_+ x_-} \over
{2 l_+ l_- -\underline{l}^2 }} = - \int_{-\infty}^{+\infty} {d l_+
\over { 2 \pi}} {p_+ (p_+ - l_+) \over {m^2}} {e^{i l_+ x_-} \over
{l_+^2 + {p_+^2 \over {m^2} }\underline{l}^2 }} \propto e^{ - x_- {
p_+ |\underline{l}| \over {m}} },
\label{yukawa}
\end{eqnarray}
which is very small. This is due to the fact that in any frame the
longitudinal separation of the nucleons ($x_-$) is much greater than
the longitudinal size of the nucleons. The non-zero mass of the quark
is not crucial, we can get the same result using some non-vanishing
quark transverse momentum $\underline{p}$ instead of the mass.

   Summing up the contributions:
\begin{eqnarray*}
A_3 + B_3 + C_3 = - i g^3 \sum_{b,c=1}^8 f^{abc} (T_1^b)(T_2^c) \left(
{k_\sigma^\perp + l_\sigma^\perp \over
{\underline{k}^2(\underline{k}+\underline{l})^2}} {1 \over {k_+ - i
\epsilon}} {1 \over {k_+ + l_+ - i \epsilon}} + {k_\sigma^\perp +
l_\sigma^\perp \over {\underline{k}^2 \underline{l}^2}} {1 \over {l_+
- i \epsilon}}{1 \over {k_+ + l_+ - i \epsilon}} \right.
\end{eqnarray*}
\begin{eqnarray}
\left. - {l_\sigma^\perp \over {\underline{k}^2 \underline{l}^2}} {1
\over {k_+ - i \epsilon}}{1 \over {l_+ - i \epsilon}} \right)(2 \pi)^2
\delta (k_-) \delta (l_-).
\end{eqnarray}
If we perform a Fourier transform of this expression and impose
$x_{-2} > x_{-1} $ condition we obtain
\begin{eqnarray}
A_3 + B_3 + C_3 = - { g^3 \over {(2 \pi )^2}} \sum_{b,c=1}^8 f^{abc}
(T_1^b) (T_2^c) \ln ( |{\underline{x}}-{\underline{x}}_1 | \lambda
){{\underline{x}}-{\underline{x}}_2 \over
|{\underline{x}}-{\underline{x}}_2|^2} \theta (x_{-} - x_{-2}),
\label{abc}
\end{eqnarray}
where $\lambda$ is some infrared cutoff, coming from the Fourier
transform of $1/{\underline{k}^2} $:
\begin{eqnarray*}
\int {d^2 \underline{k} \over { (2 \pi)^2}} e^{ - i \underline{k}
 \cdot {\underline{x}}} {1 \over {\underline{k}^2}} = - {1 \over {2
 \pi}} \ln (|\underline{x}| \lambda).
\end{eqnarray*}
Now our claim becomes manifest. Summing the expressions like
Eq. (\ref{abc}) for different pairs of quark lines we see that the
cutoff $\lambda$ gets cancelled, and we end up with an expression
exactly equal to the one given in Eq. (\ref{f3}).

\begin{figure}
\begin{center}
\epsfxsize=10cm
\epsfysize=18cm
\leavevmode
\hbox{ \epsffile{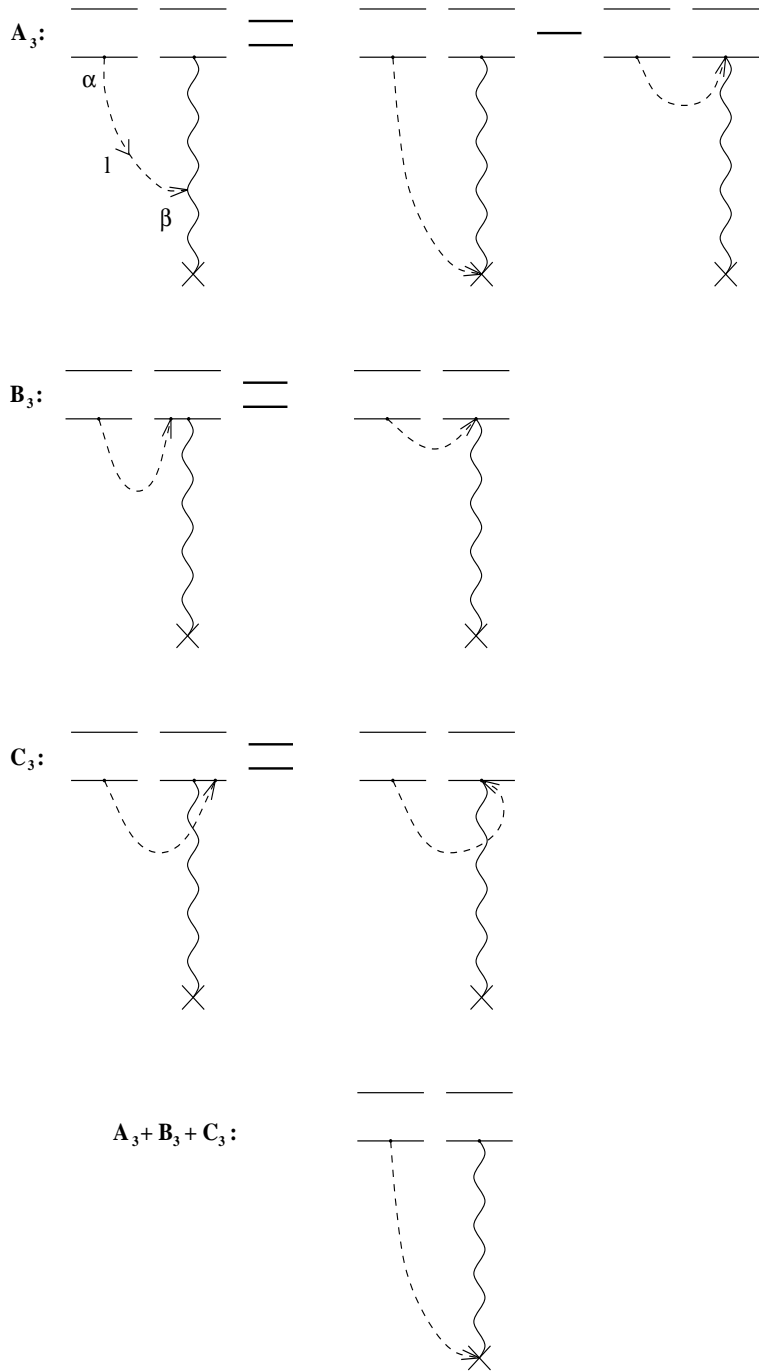}}
\end{center}
\caption{ Application of the Ward identity at order $g^3$.}
\label{ward}
\end{figure}

     The principle behind this summation of diagrams is the Ward
identity. The covariant part of the propagator of the $l$-line in the
graph $A_3$ in Fig. \ref{or3} in coordinate space gives a contribution
proportional to $\theta (x_{-1} - x_{-2})$, which is excluded by our
ordering of the nucleons: $x_{-2} > x_{-1}$. One can track this
explicitly through the calculations, or use the following
``heuristic'' argument. If we have only the covariant part of the
$l$-line propagator, then the three-gluon vertex in the graph $A_3$
(Fig. \ref{or3}) should be close to the first (left) nucleon in the
longitudinal direction, since the covariant part of the gluon
propagator can not propagate over large distances along the $x_-$-axis
[see Eq. (\ref{yukawa})]. Then the $k$-line should propagate the
distance between the two nucleons, so that its propagator can not have
a covariant part.  But, because of the current conservation this
propagator contains only a $k_\mu \eta_\nu \over {k_+ - i \epsilon}$
term and, therefore, can not go backwards in the $x_-$-direction. So,
once we impose the ordering of the nucleons along the $x_-$-axis this
contribution becomes zero. It was shown above that the contribution of
the covariant part of the $l$-line propagator is also zero for the
graphs $B_3$ and $C_3$ in Fig. \ref{or3}. We can conclude that the
$l$-line is longitudinally polarized at its right end in all of the
three graphs in Fig. \ref{or3}, and, consequently, we can apply Ward
identity.

    The way to apply it at order $g^3$ is illustrated in
Fig. \ref{ward}. We follow the notation introduced by t'Hooft in
\cite{t'Hooft}, which is also described in \cite{Sterman}. The dashed
line in Fig. \ref{ward} corresponds to a longitudinally polarized
gluon. The propagator for this line is $ - {i \over {\underline{l}^2}}
{\eta_\alpha l_\beta \over {l_{+} - i \epsilon}}$, where the arrow
corresponds to the $\beta$-end of the line. The beginning of the line
($\alpha$-end ) is just a usual QCD vertex, in our case the
gluon-fermion vertex. On the left-hand side of Fig. \ref{ward} the
vertex at the other end of the line, where the arrow is, is also a QCD
vertex. However, on the right-hand side of Fig. \ref{ward} it implies
only the four-momentum conservation and gives no other factors. The
color factors of the graphs on the right-hand side of Fig. \ref{ward}
are the same as the color factors on the left-hand side. After we
apply the Ward identity we get the contributions on the right hand
side. The graphs where the dashed line hooks to the end of a quark
line are zero, since the quarks are on-shell. That is why we do not
have such contributions for $B_3$ and $C_3$.  The diagrams on the
right hand side of $B_3$ and $C_3$ cancel the second diagram on the
right-hand side of the expression for $A_3$. We are left with the
first diagram, which gives the answer (see Fig. \ref{ward}).

    So far we have calculated only the $\sigma = \perp$ component of
the diagrams on Fig. \ref{or3}. To get a full correspondence to the
classical field one needs to show that $\sigma = +$ and $\sigma = -$
contributions are zero. From the light-cone gluon propagator we
obviously see that $\sigma = +$ component is zero. To get the $\sigma
= -$ component one has to take the $\eta_\sigma (l+k)_\rho \over {l_+
+ k_+ + i\epsilon}$ term in the propagator, which is longitudinally
polarized at the $\rho$-end. Summing over all possible connections of
this line to the gauge invariant object above (two nucleons connected
by a gluon line) we get zero due to the Ward identity. Note that these
connections include some diagrams which are not shown in
Fig. \ref{or3}, since they give obviously wrong $x_-$ ordering in
coordinate space.

\section{Higher orders}  

     Here we are going to work with those diagrams giving the
classical field at order $g^5$. We first note that we are looking for
a correspondence between the diagrams and the classical gluon field
taken in the form in which it appears in the gluon distribution
function, i.e., in the correlation function of two classical
fields. But when we calculate a correlation function, we have to
impose the condition that each nucleon is a color singlet, and average
over all possible colors (see \cite{me,Larry,Alex,Jamal}). In the
spirit of the calculation of the gluon distribution function, we will
treat the first nucleon as a color singlet, which means that we will
take a trace in this nucleon's color space. We will do this for the
diagrams, as well as for the classical solution itself. Then the color
averaged, in the color space of the first nucleon, classical solution
at order $g^5$ is
\begin{eqnarray}
\left<{\underline {A}}^a (\underline {x}, x_{-} ) |_{o(g^5)}\right>_1
= - { g^5 \over {4 (2 \pi)^3}} (T^a_2) \ln^2 \left(
{|{\underline{x}}-{\underline{x}}_1 | \over
|{\underline{x}}-{\underline{x}'}_1|} \right) \left(
{{\underline{x}}-{\underline{x}}_2 \over
|{\underline{x}}-{\underline{x}}_2|^2}\theta (x_{-} - x_{-2}) -
{{\underline{x}}-{\underline{x}'}_2 \over
|{\underline{x}}-{\underline{x}'}_i|^2}\theta (x_{-} -
x'_{-2})\right).
\label{f5}
\end{eqnarray}

\begin{figure}
\begin{center}
\epsfxsize=12cm 
\epsfysize=18cm 
\leavevmode 
\hbox{ \epsffile{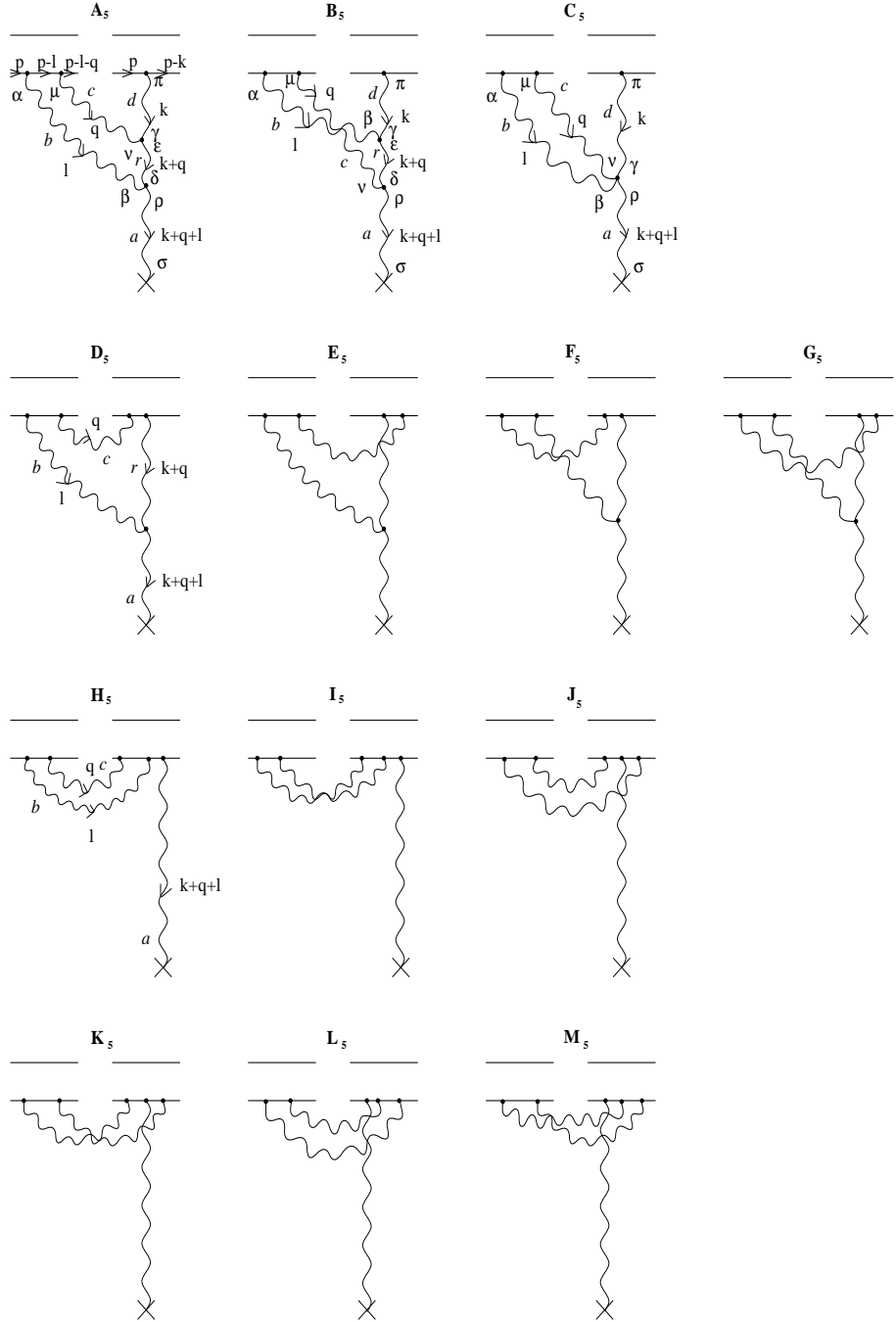}}
\end{center}
\caption{ The diagrams at order $g^5$.}
\label{or5}
\end{figure}

    Let us calculate the contributions of the graphs shown in
Fig. \ref{or5}, doing the color averaging mentioned above. The lines
connected to the first nucleon will always have momenta $l$ and $q$,
the line connected to the second nucleon will carry momentum $k$, just
as in graph $A_5$ in Fig. \ref{or5}. We will keep the parts of the
$l$- and $q$- lines' propagators which are longitudinally polarized at
the right end, i.e., the ${\eta_\alpha l_\beta \over {l_{+} - i
\epsilon}}$ and ${\eta_\mu q_\nu \over {q_{+} - i \epsilon}}$
parts. The contributions where at least one of these lines is
longitudinally polarized at the opposite end will vanish after
applying the Ward identity and color averaging in the color space of
the first (left) nucleon. So, we throw away those parts of the
propagators.  The contributions where we take one or both of $l$- and
$q$- lines to be covariant give us the terms proportional to $\theta
(x_{-1} - x_{-2})$, which is zero. This can be shown by a brute force
calculation or by a ``heuristic'' argument, similar to the one given
at order $g^3$. Finally we are left with the ${\eta_\alpha l_\beta
\over {l_{+} - i \epsilon}}$ and ${\eta_\mu q_\nu \over {q_{+} - i
\epsilon}}$ parts of the propagators, which give us some non-vanishing
contribution.

    At order $g^5$ the $\sigma = +$ and $\sigma = -$ components of the
diagrams are zero, for the same reasons as at the order $g^3$. When we
apply Ward identity to get the cancelation of $\sigma = -$ component,
similarly to $o(g^3)$ case, we have to include several diagrams which
are not present in Fig. \ref{or5}, but go away after color averaging
in the first nucleon or because they have a wrong $x_-$ ordering in
coordinate space. Some of these diagrams are divergent, i.e., purely
quantum, but those vanish after color averaging.  Therefore we
concentrate our efforts on $\sigma = \perp$ part. Since both $l$- and
$q$- lines are longitudinally polarized we can apply Ward identity to
sum these diagrams. That is we can draw a bunch of pictures like those
in Fig. \ref{ward}, get some cancelations, and end up with the
answer. In the spirit of this approach we regroup the terms in the
contributions of each diagram (before doing the color averaging) in
the following way, and where a summation over the repeating indices is
assumed).

\begin{mathletters}
\begin{eqnarray*}
A_5 = g^5 f^{arb} f^{rcd} (T^d_2) (T^c_1 T^b_1) { 1 \over {
{\underline{l}}^2 {\underline{q}}^2}} { 1 \over {l_{+} - i \epsilon}}
{ 1 \over {q_{+} - i \epsilon}} \left( {k_\sigma^\perp \over
{{\underline{k}}^2}} { 1 \over {k_{+} - i \epsilon}} -
{(k+q)_\sigma^\perp \over {({\underline{k}} +{\underline{q}})^2}} { 1
\over {k_+ + q_+ - i \epsilon}} \right.
\end{eqnarray*}
\begin{eqnarray}
\left. - {1 \over {2 p_+}} \tilde{u} (p-k) \gamma_\pi u(p) q_\nu
P_{\pi \gamma} (k) \Gamma_{\nu \epsilon \gamma} P_{\epsilon \sigma}
(k+q+l) \right) \pi (2 \pi)^2 \delta (k_-) \delta (l_-) \delta (q_-),
\end{eqnarray}
\begin{eqnarray*}
B_5 = g^5 f^{arc} f^{rbd} (T^d_2) (T^c_1 T^b_1){ 1 \over {
{\underline{l}}^2 {\underline{q}}^2}} { 1 \over {l_{+} - i \epsilon}}
{ 1 \over {q_{+} - i \epsilon}} \left( - {(k+l)_\sigma^\perp \over
{({\underline{k}} +{\underline{l}})^2}} { 1 \over {k_+ + l_+ - i
\epsilon}} +{(k+l+q)_\sigma^\perp \over {({\underline{k}}
+{\underline{l}} +{\underline{q}} )^2}} { 1 \over {k_+ + l_+ + q_+ - i
\epsilon}} \right.
\end{eqnarray*}
\begin{eqnarray}
\left. + {1 \over {2 p_+}} \tilde{u} (p-k) \gamma_\pi u(p) q_\nu
P_{\pi \delta} (k) \Gamma_{\rho \nu \delta} P_{\rho \sigma} (k+q+l)
\right) \pi (2 \pi)^2 \delta (k_-) \delta (l_-) \delta (q_-),
\end{eqnarray}

\begin{eqnarray*}
C_5 = g^5 (T^d_2) (T^c_1 T^b_1){ 1 \over { {\underline{l}}^2
{\underline{q}}^2}} { 1 \over {l_{+} - i \epsilon}} { 1 \over {q_{+} -
i \epsilon}}{1 \over {2 p_+}} \tilde{u} (p-k) \gamma_\pi u(p) q_\nu
l_\beta P_{\pi \gamma} (k) \Gamma_{\rho \beta \nu \gamma}^{abcd}
P_{\rho \sigma} (k+q+l)
\end{eqnarray*}
\begin{eqnarray}
\times \pi (2 \pi)^2 \delta (k_-) \delta (l_-) \delta (q_-) ,
\end{eqnarray}

\begin{eqnarray*}
D_5 + E_5 = g^5 f^{arb} f^{rcd} (T^d_2) (T^c_1 T^b_1) { 1 \over {
{\underline{l}}^2 {\underline{q}}^2}} { 1 \over {l_{+} - i \epsilon}}
{ 1 \over {q_{+} - i \epsilon}} \left( {(k+q)_\sigma^\perp \over
{({\underline{k}} +{\underline{q}})^2}} { 1 \over {k_+ + q_+ - i
\epsilon}} \right.
\end{eqnarray*}
\begin{eqnarray}
\left.  - {(k+l+q)_\sigma^\perp \over {({\underline{k}}
+{\underline{l}} +{\underline{q}} )^2}} { 1 \over {k_+ + l_+ + q_+ - i
\epsilon}} \right) \pi (2 \pi)^2 \delta (k_-) \delta (l_-) \delta
(q_-) ,
\end{eqnarray}

\begin{eqnarray*}
F_5 + G_5 = g^5 f^{arc} f^{rbd} (T^d_2) (T^c_1 T^b_1){ 1 \over {
{\underline{l}}^2 {\underline{q}}^2}} { 1 \over {l_{+} - i \epsilon}}
{ 1 \over {q_{+} - i \epsilon}} \left({(k+l)_\sigma^\perp \over
{({\underline{k}} +{\underline{l}})^2}} { 1 \over {k_+ + l_+ - i
\epsilon}} \right.
\end{eqnarray*}
\begin{eqnarray}
\left. - {(k+l+q)_\sigma^\perp \over {({\underline{k}}
+{\underline{l}} +{\underline{q}} )^2}} { 1 \over {k_+ + l_+ + q_+ - i
\epsilon}} \right) \pi (2 \pi)^2 \delta (k_-) \delta (l_-) \delta
(q_-),
\end{eqnarray}
\begin{eqnarray*}
H_5 + I_5 + J_5 + K_5 + L_5 + M_5 = g^5 (T^c_1 T^b_1)
{(k+l+q)_\sigma^\perp \over { {\underline{l}}^2 {\underline{q}}^2
({\underline{k}} +{\underline{l}} +{\underline{q}} )^2}} { 1 \over
{l_{+} - i \epsilon}} { 1 \over {q_{+} - i \epsilon}}{ 1 \over {k_+ +
l_+ + q_+ - i \epsilon}} \left[ -f^{abr} f^{rcd} (T^d_2) \right.
\end{eqnarray*}
\begin{eqnarray*}
\left. + {1 \over {2 p_+}} \tilde{u} (p-k) \left( \gamma_+ {1 \over
{\gamma \cdot (p+l+q)}} \gamma \cdot q (T^a_2 [T^c_2, T^b_2]) + \gamma
\cdot q {1 \over {\gamma \cdot (p-k-l-q)}} \gamma_+ ([T^c_2, T^b_2]
T^a_2) \right) u(p) \right]
\end{eqnarray*}
\begin{eqnarray}
\times \pi (2 \pi)^2 \delta (k_-) \delta (l_-) \delta (q_-) ,
\end{eqnarray}
\end{mathletters}
where $\Gamma_{\nu \epsilon \gamma}$ is the three gluon vertex,
omitting color dependence, $\Gamma_{\rho \beta \nu \gamma}^{abcd}$ is
a four-gluon vertex including the color factors, and $P_{\alpha \beta}
(k)$ is the gluon's propagator.

    By writing the delta functions of the minus components of the
momenta we are already anticipating the color averaging. After taking
the trace in the color space of the first nucleon everything becomes
symmetric under the $l \leftrightarrow q$ interchange. The quark line
in the first nucleon for any graph in Fig. \ref{or5} gives ${1 \over
{2 p_+}} \tilde{u} (p-l-q) \gamma_+ { (p-l) \cdot \gamma \over {
(p-l)^2 + i \epsilon}}\gamma_+ u(p) (2 \pi) \delta(l_- + q_-) \approx
- {1 \over {l_- - i \epsilon}} (2 \pi) \delta(l_- + q_-)$. Using the
$l \leftrightarrow q$ symmetry we can symmetrize this result, by just
switching $l$- and $q$- lines. We obtain: ${1 \over {2}} \left( - {1
\over {l_- - i \epsilon}} - {1 \over {q_- - i \epsilon}} \right) (2
\pi) \delta(l_- + q_-) = {1 \over {2}} \left( {1 \over {l_- + i
\epsilon}} - {1 \over {l_- - i \epsilon}} \right) (2 \pi) \delta(l_- +
q_-) = - \pi i \delta(l_-) (2 \pi) \delta(q_-)$, which we included in
the contributions of the diagrams. When the $l$- and $q$- gluons hook
to different quark lines in the nucleon, we get the similar factors
even without color averaging.

     After summing all the contributions and taking the color average,
and after some algebra which we are going to skip, we end up with
\begin{eqnarray}
\left<A_5 + \ldots +M_5 \right>_1 = g^5 (T^a_2) { k_\sigma^\perp \over
{{\underline{k}}^2 {\underline{l}}^2 {\underline{q}}^2}} { 1 \over
{k_{+} - i \epsilon}} { 1 \over {l_{+} - i \epsilon}} { 1 \over {q_{+}
- i \epsilon}} \pi^2 (2 \pi) \delta (k_-) \delta (l_-) \delta (q_-),
\end{eqnarray}
which, after a Fourier transform, gives
\begin{eqnarray}
\left<A_5 + \ldots +M_5 \right>_1 = - { g^5 \over {4 (2 \pi)^3}}
(T^a_2) \ln^2 \left( |{\underline{x}}-{\underline{x}}_1 | \lambda
\right) {{\underline{x}}-{\underline{x}}_2 \over
|{\underline{x}}-{\underline{x}}_2|^2}\theta (x_{-} - x_{-2}).
\label{d5}
\end{eqnarray}
Now it becomes obvious that after summing over all possible
connections to the quark lines of the $l$- and $q$- gluons in the
first nucleon and of $k$- gluon in the second we will reproduce
formula (\ref{f5}). The color averaging is crucial, it eliminates many
extra terms in the sum of the contributions of different diagrams. It
also eliminates some graphs at order $g^5$ which have ``quantum''
parts --- vertex and propagator corrections. If we had not imposed the
color singlet condition, the correspondence between the classical
field and the diagrams would not work.

\begin{figure}
\begin{center}
\epsfxsize=8cm 
\epsfysize=5cm 
\leavevmode 
\hbox{ \epsffile{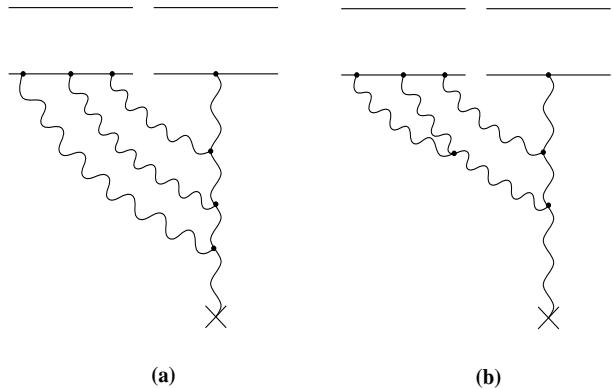}}
\end{center}
\caption{(a) A typical ``classical'' diagram at the order $g^7$, (b) a
diagram which is not included in the classical field at order $g^7$,
but doesn't vanish.}
\label{or7}
\end{figure}

     One can ask the question whether it is possible to go to the
higher orders in $g$ that is, to orders $g^7$, $g^9$, etc. The answer
is no, because at higher orders the classical field does not dominate,
and the contribution of quantum corrections becomes important. We
illustrate this statement at order $g^7$ in Fig. \ref{or7}. The
diagram in Fig. \ref{or7}(a) is a typical graph which one would expect
to contribute to the classical gluon field at this
order. Fig. \ref{or7}(b) is one of the many divergent diagrams for the
gluon field at the order $g^7$. Here one can not eliminate the
``quantum'' graph given in Fig. \ref{or7}(b) by color averaging as was
done at lower orders. There is no other reason for this graph to be
suppressed.  Therefore, both of the graphs in Fig. \ref{or7}
contribute at this order.
The diagram in Fig. \ref{or7}(b) can not be a part of the classical
field, because it is divergent and has to be renormalized, which is an
essentially quantum procedure. So, the gluon field at this order has
both classical and quantum contributions in it, and, although the
correspondence of the classical field to some diagrams may still hold,
it doesn't make much physical sense to try to isolate it. Therefore
once we have more than two gluons connected to the first nucleon we
can not take the gluon field to be classical.  This way we obtained a
limit to McLerran-Venugopalan model. The classical approach is valid
as long as we have no more than two gluons per nucleon.

\section{Conclusions}

\begin{figure}
\begin{center}
\epsfxsize=10cm 
\epsfysize=4cm 
\leavevmode 
\hbox{ \epsffile{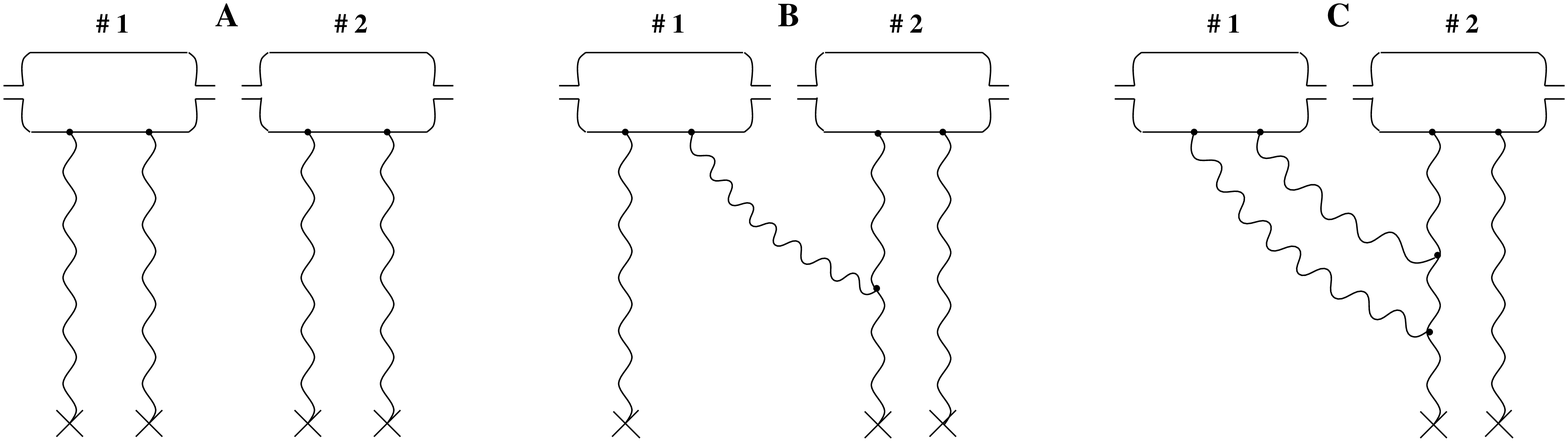}}
\end{center}
\caption{Diagrams contributing to the scattering amplitude at order
$g^8$ for two nucleons (see text).}
\label{ampl}
\end{figure}

    To illustrate this limit, still at the level of two interacting
nucleons, we will construct diagrams contributing to the cross-section
of the nucleus on a quarkonium (quark-antiquark bound state) at order
$g^8$, which means two gluons per nucleon. An important parameter in
McLerran-Venugopalan model is $N {\alpha_s}^2$, where $N$ is the
number of nucleons in the nucleus. It plays the role of an effective
coupling.  The kinematic region we are considering is $N {\alpha_s}^2
\sim 1$, $\alpha_s \ll 1$. In the process we are going to consider
there will be only two participating nucleons. (There are $N$ nucleons
in the nucleus but, for simplicity, we allow only two of them
interact.). Then, in terms of that ``effective coupling'' of the
theory, the process will be of the order $(N {\alpha_s}^2)^2$, which
exactly corresponds to order $g^8$ diagrams. The diagrams that survive
are shown in Fig. \ref{ampl}. The quark lines of the onium (not shown
in Fig. \ref{ampl}) connect to the crosses at the ends of the gluon
lines. Each cross represents a gluon field.  The generalization to
more than two interacting nucleons is straightforward.

By diagram $B$ in Fig. \ref{ampl} we mean a class of diagrams where
the gluon line coming from the first (left) nucleon to the second
nucleon connects in all possible ways to the second nucleon and the
gluons emitted off it. Similarly graph $C$ in Fig. \ref{ampl} includes
all diagrams where the two gluon lines connecting the nucleons hook in
all possible ways to the second nucleon. Also it is understood that in
all graphs gluons hook to all possible quark lines in the
nucleons. Color averaging is assumed in the color space of each of the
nucleons.

    We have to prove that the graphs we drew are the only possible
diagrams giving significant contribution to the scattering.  We do not
consider the diagrams where all gluons hook to one of the nucleons and
the other nucleon remains a non-interacting spectator or just
interacts with itself. Those graphs would be at most of order $N
{\alpha_s}^4$, i.e., down by a factor of $N$ compared to the diagrams
in Fig. \ref{ampl}. One can easily see that for a graph corresponding
to four gluon fields coming off two nucleons, the diagrams of type $A$
are the only possibilities at order $g^8$. (Note that now we do the
color averaging everywhere.) Analogously we can prove that graphs like
$B$ are the only possibilities for three-field contributions at this
order of the coupling. The fourth gluon line can not remain in just
one nucleon since that contribution will be cancelled by color
averaging. It has to connect to another nucleon and that way we obtain
graphs like $B$. ``Symmetric'' graphs, i.e., the graphs where the
first and the second nucleon are interchanged in the diagram, but not
in the $x_-$ direction, are either equivalent to the diagrams in class
$B$, as happens to that particular graph shown in Fig. \ref{ampl}, or
give zero after imposing longitudinal ordering and applying Ward
identities.  The arguments leading to this conclusion are much the
same as those we now give for graphs in class $C$.

\begin{figure}
\begin{center}
\epsfxsize=8cm 
\epsfysize=10cm 
\leavevmode 
\hbox{ \epsffile{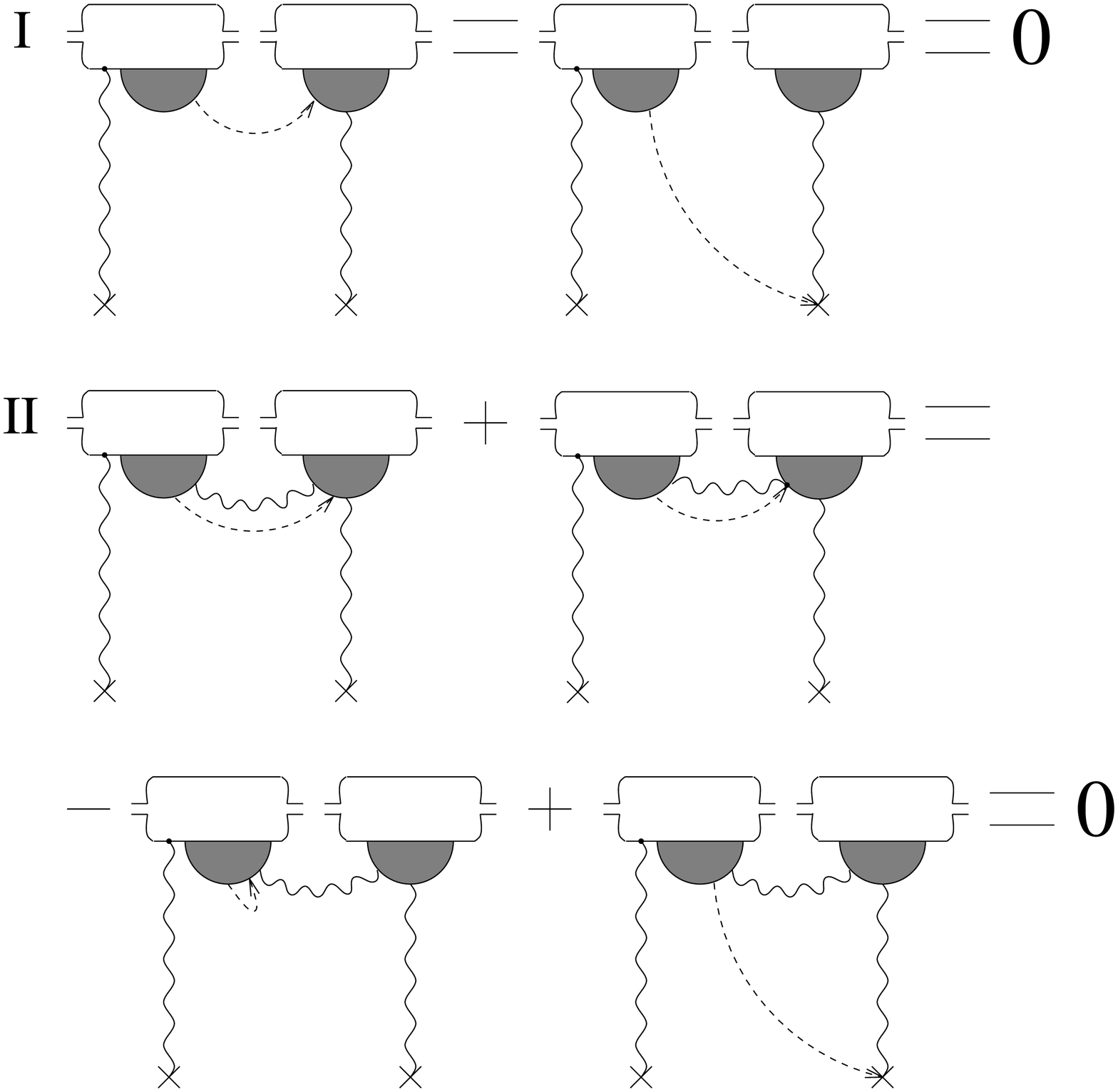}}
\end{center}
\caption{The way to eliminate many of the diagrams, which could appear
in the nucleus-onium scattering. }
\label{zero}
\end{figure}

    For the graphs with two gluon fields the situation is a little
more complicated. Here we have many more diagrams which disappear
leaving only diagrams as in $C$ in Fig. \ref{ampl}. Most of the graphs
with two gluon fields at order $g^8$, which are not equivalent to
diagrams in class $C$, can be represented as having one gluon line,
which gives the field, connected to some fermion line in nucleon
number one, with the other three lines connected in all possible ways
to provide one more gluon field, but not hooking to that first gluon
line. Now, in each of these diagrams there must be one or two paths to
get from one nucleon to the other along gluon lines. Each of these
paths corresponds to some product of the propagator denominators. Due
to the longitudinal separation between the nucleons , at least one of
these denominators should include $1 \over {l_+ - i \epsilon}$. That
is, it should correspond to a part of propagator which is
longitudinally polarized at the right end, ${ \eta_\alpha l_\beta
\over {l_+ - i \epsilon}} $, otherwise we would get either an
exponential suppression as in (\ref{yukawa}) or a wrong ordering. Such
a denominator should be present on each path from one nucleon to the
other. This is illustrated in Fig. \ref{zero}. The dashed line
corresponds to the longitudinally polarized propagator, and is the
same as in Fig. \ref{ward}. The blobs represent some combinations of
the gluon lines. Each graph is of order $g^8$. Case I corresponds to
diagrams where there is only one path from one nucleon to the other
along the gluon lines in the diagram. It also includes the case where
there are two paths, but a longitudinally polarized line belongs to
both of them. The situation where we have two paths between the
nucleons and the longitudinally polarized lines are different for each
of the paths is represented in case II. There we pick the
longitudinally polarized line along one of the paths, without worrying
much about the location of the similar line on the other path.

   For case I in Fig. \ref{zero} we can just apply the Ward identity
to get the diagram on the right, which is zero after color averaging
in the right nucleon.  By application of the Ward identity we mean
summation over the contributions of the diagrams which have the same
structure to the left and to the right of the dashed line, but
different connections of the dashed line on the right hand side. The
sum of these gives the graph on the right of I, similarly to
Fig. \ref{ward}. Analogously in case II in Fig. \ref{zero} we can
apply the Ward identity to the dashed line. The result is shown in the
second line of case II in Fig. \ref{zero}. The first graph there
corresponds to the arrow of the dashed line connected either to the
vertex where the two paths split, if such vertex exists, or to the
quark line in the first nucleon, if the paths do not overlap in the
left blob. Now in both graphs there is only one path from one nucleon
to another, and that brings us back to case I and cancels for the same
reason. That way we include all the contributions from these diagrams
and prove that they are zero.

\begin{figure}
\begin{center}
\epsfxsize=8cm 
\epsfysize=4cm 
\leavevmode 
\hbox{ \epsffile{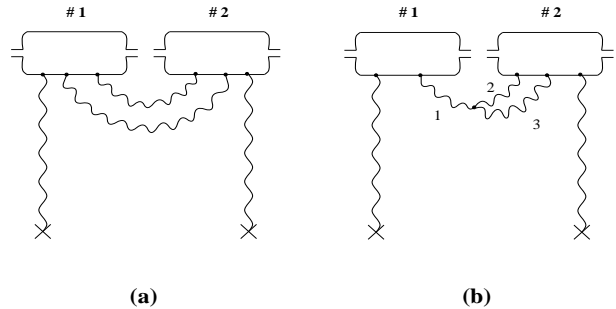}}
\end{center}
\caption{Examples of the graphs that vanish (see text).}
\label{sample}
\end{figure}

    We illustrate our technique in Fig. \ref{sample}. In general at
order $g^8$ the graphs that we consider in Fig. \ref{zero} can be
subdivided in two classes, representatives of which are shown in
Fig. \ref{sample}. We may have two gluons leaving nucleon number one
to connect to nucleon number two [Fig. \ref{sample}(a)]. There may
also be just one such gluon [Fig. \ref{sample}(b)]. The application of
our method to the class of diagrams in Fig. \ref{sample}(a) is
straightforward. This obviously corresponds to case II in
Fig. \ref{zero}. At least one of the gluon lines connecting the
nucleons should be longitudinally polarized. Summation over all of its
possible connections on the right and application of the Ward identity
leaves us with only one line connecting the nucleons. This line in its
turn should be longitudinally polarized. Applying the Ward identity
once again we get zero.

    The diagrams represented in Fig. \ref{sample}(b) are a little
harder to deal with. The contribution in which line 1 is
longitudinally polarized on the right belongs to case I in
Fig. \ref{zero}. Therefore, summing over all possible connections of
line 1 on the right we get zero. When line 1 is covariant or
longitudinally polarized at the left end we need either line 2 or line
3 to be longitudinally polarized at the right end. Let it be line
3. Now the situation corresponds to case II in
Fig. \ref{zero}. Applying the Ward identity we end up with the right
end of line 3 hooking back to the three-gluon vertex or to the cross
at the end of the gluon connected to the second nucleon. We took the
contribution of line 1 which can not insure the longitudinal
separation. Therefore, now line 2 has to be longitudinally polarized
on the right. The situation again becomes similar to case I in
Fig. \ref{zero}. Summation over all possible connections of line 2 on
the right gives us zero.

    The diagrams which are not included in the representation shown in
Fig. \ref{zero} can be eliminated by a similar technique. In the end
we are left with the class of the diagrams $C$ in Fig. \ref{ampl}. The
classical field at order $g^5$ is included in contributions of some of
these diagrams. The initial and final states of the nucleons are color
singlets. Therefore, color averaging in the first nucleon when
calculating the graphs for the classical field at order $g^5$ in
Sect. III is justified. There are no graphs with just one gluon field
contributing to the scattering process, since we can not emit one
gluon off a color singlet object.

    So, the scattering process in light-cone gauge is described by the
diagrams of the types shown in Fig. \ref{ampl}, i.e., by the classical
field. If one thinks about this process in the covariant gauge, it is
easy to see, that the only diagrams that contribute are of type $A$ in
Fig. \ref{ampl}. It obviously is a combination of the classical
fields. That way the correspondence can be easily seen in the
covariant gauge. The two gluons per nucleon limit is also manifest in
that gauge.

    To conclude we summarize the results of this paper. The Feynman
diagrams corresponding to the classical non-Abelian
Weizs\"{a}cker-Williams field in the light-cone gauge were constructed
(see Fig. \ref{or1}, Fig. \ref{or3}, Fig. \ref{or5} ). We derived a
limit for the classical approach, which is two gluons per nucleon.  It
was shown that for a large nucleus the diagrams, satisfying this
limit, which dominate a scattering process are described by a
classical field (see Fig. \ref{ampl}). Therefore it is possible that
the classical field may be used for calculation of such processes as
charm production, dijet cross-section and many others in nuclear
collisions.

\section*{Acknowledgments}
   
     I wish to thank Professor A. H. Mueller for suggesting this work
and for numerous stimulating and informative discussions, as well as
for reading the manuscript. I thank the Department of Energy's
Institute for Nuclear Theory at the University of Washington for its
hospitality and for partial support during the final stages of this
work. In particular I would like to thank Miklos Gyulassy, Jamal
Jalilian-Marian, Alex Kovner, Andrei Leonidov, Larry McLerran, Dirk
Rischke, and Raju Venugopalan for many interesting discussions. This
research is sponsored in part by the US Department of Energy under
grant DE-FG02 94ER 40819.

\end{document}